\documentclass[aps,twocolumn,showpacs,preprintnumbers,amsmath,amssymb,showpacs,showkeys]{revtex4}
\usepackage{slashed}
\usepackage{amssymb}
\usepackage{amsmath}
\usepackage{amsfonts}
\usepackage{epsfig}
\usepackage{graphicx}
\usepackage{tabularx}
\newcommand{\seq}{\begin{subequations}}
\newcommand{\sen}{\end{subequations}}
\newcommand{\eq}{\begin{eqnarray}}
\newcommand{\en}{\end{eqnarray}}

\def\shiftdown#1{#1\llap{\lower.04ex\hbox{#1}}}

\begin{document}

\title{Analysis of soft wall AdS / QCD potentials to obtain melting temperature of scalar hadrons}

\author{Alfredo Vega and Adolfo Iba\~nez}
\vspace*{1.2\baselineskip}

\affiliation{
Instituto de F\'isica y Astronom\'ia y Centro de Astrof\'isica de Valpara\'iso, 
     Universidad de Valpara\'iso,\\
     Avenida Gran Breta\~na 1111, Valpara\'iso, Chile
\vspace*{1.2\baselineskip}\\}


\begin{abstract}

We consider an analysis of potentials related to Schr\"odinger-type equations for scalar fields in a 5D AdS black hole background with dilaton in order to get melting temperatures for different hadrons in a thermal bath. The approach does not consider calculations of spectral functions and it is easy to get results for hadrons with an arbitrary number of constituents. We present results for scalar mesons, glueballs, hybrid mesons and tetraquarks, and we show  that mesons are more resistant to being melted in a thermal bath than other scalar hadrons, and in general the melting temperature increases when hadrons contain heavy quarks.

\end{abstract}

\date{\today}



\maketitle

\section{Introduction}

Today in several places Heavy Ion Collision experiments are being conducted to study the quark gluon plasma (QGP), and in this way understand how temperature and high density media can affect hadron properties. This topic has attracted the interest of theoretical physicists, who have considered several approaches to describe hadrons under these conditions (e.g., see \cite{Gyulassy:2004zy, Shuryak:2004cy}). The main theoretical tools used in this area are for example lattice QCD  sum rules, and in last 15 years techniques based on  Gauge / Gravity dualities have been added (e.g., see \cite{Erdmenger:2007cm, Kajantie:2006hv, Colangelo:2013ila, Colangelo:2012jy, Braga:2016wkm}). In the latter kind of models,  temperature is included, considering black hole backgrounds in AdS in $d+1$ dimensions.

One of the interesting properties for hadrons in a high temperature environment is their melting temperature, because there are reasons to believe that different hadron species are melted at different temperatures, and some of them  can resist temperatures higher than necessary to start to form a QGP \cite{Matsui:1986dk, Karsch:1987pv}, and the suppression of these species can be used as a signal to know if this plasma was created and give an idea of the temperature reached.

One way to obtain the melting temperature is by calculating the spectral function and studying temperature where the maximum dissapears. This strategy has been followed in different theoretical approaches (e.g., \cite{Ayala:2016vnt}), as in several works which use Gauge / Gravity ideas (e.g., see \cite{Colangelo:2013ila, Colangelo:2012jy, Braga:2016wkm}).

The holographic correspondence, in particular in the soft wall approach, offers an alternative to calculate melting temperatures for hadrons in a thermal bath without calculating the spectral function; it is only necessary to study the potential related to the Schr\"dinger-type equation that describes hadrons on the AdS side. According to the literature, thereare two approaches that consider an analysis of potential to get hadron melting temperatures, and both are based on the possibility of performing a transformation in the equation of motion that describes hadron modes in an AdS black hole background to obtain a Schr\"odinger-type equation. One of them considers a simple transformation to get a potential that depends on temperature and hadron mass also \cite{Fujita:2009ca}; therefore, it is necessary to have a prior knowledge of hadron mass or to make a supposition (in general authors use mass values at zero temperature), and in this case it is possible to obtain an approximate value for melting temperature of the last hadrons to survive in the thermal bath. The other approach considers a more elaborate transform which produces a potential independent of the hadron mass, and in this case it is not neccessary to make assumptions on hadron mass, so it produce  produce a more precise value for the hadron melting temperature \cite{Miranda:2009qp, Bellantuono:2014lra, Bartz:2016ufc}. In this work, we adopt the second approach and we also consider the relation between AdS mass and the dimension of operators according to the AdS / CFT dictionary \cite{Aharony:1999ti}; thus, the model can describe hadrons in general with an arbitrary number of constituents.

In this paper, we limit ourselves to the study of scalar hadrons and show results for melting temperatures for mesons, glueballs, hybrid mesons and tetraquarks in a thermal bath. We consider a holographic soft wall model with dilaton for light and heavy hadrons. We found in general that mesons are more resistant to being melt in a thermal bath than other scalar hadrons, such as scalar glueballs, hybrids, etc, and that melting temperature increases when considering hadrons with heavy quarks.

This paper consists of three sections and an appendix. This introduction is followed by section 2, where we describe the model and procedure used to calculate melting temperatures on the AdS side, and in section 3 we present our results and conclusions, including future perspectives. Additionally, we include an appendix with the Liouville substitution, which is the procedure used to get Schr\"odinger-type equations and potentials, which are analysed in this work to obtain melting temperatures.



\begin{center}
\begin{figure*}
  \begin{tabular}{c c}
    \includegraphics[width=3.1 in]{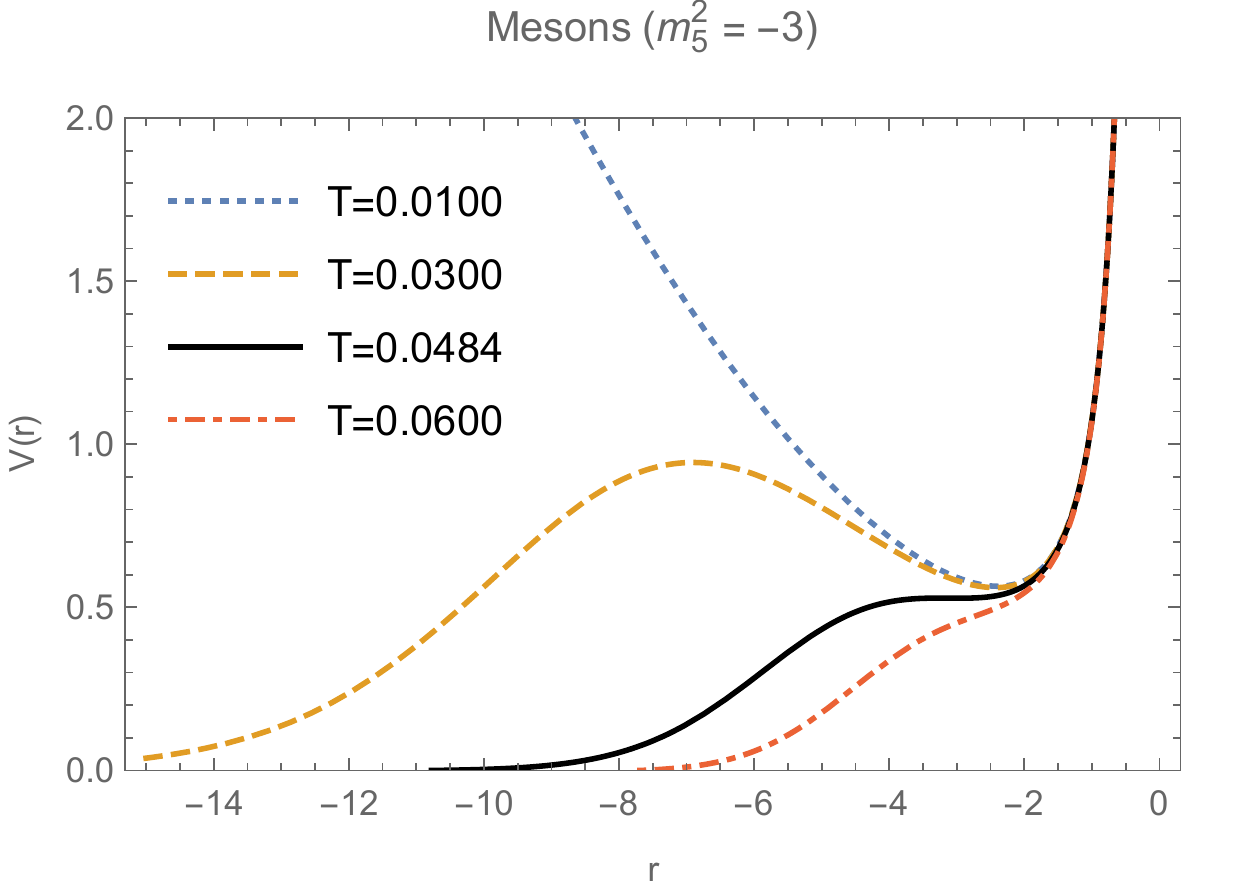}
    \includegraphics[width=3.1 in]{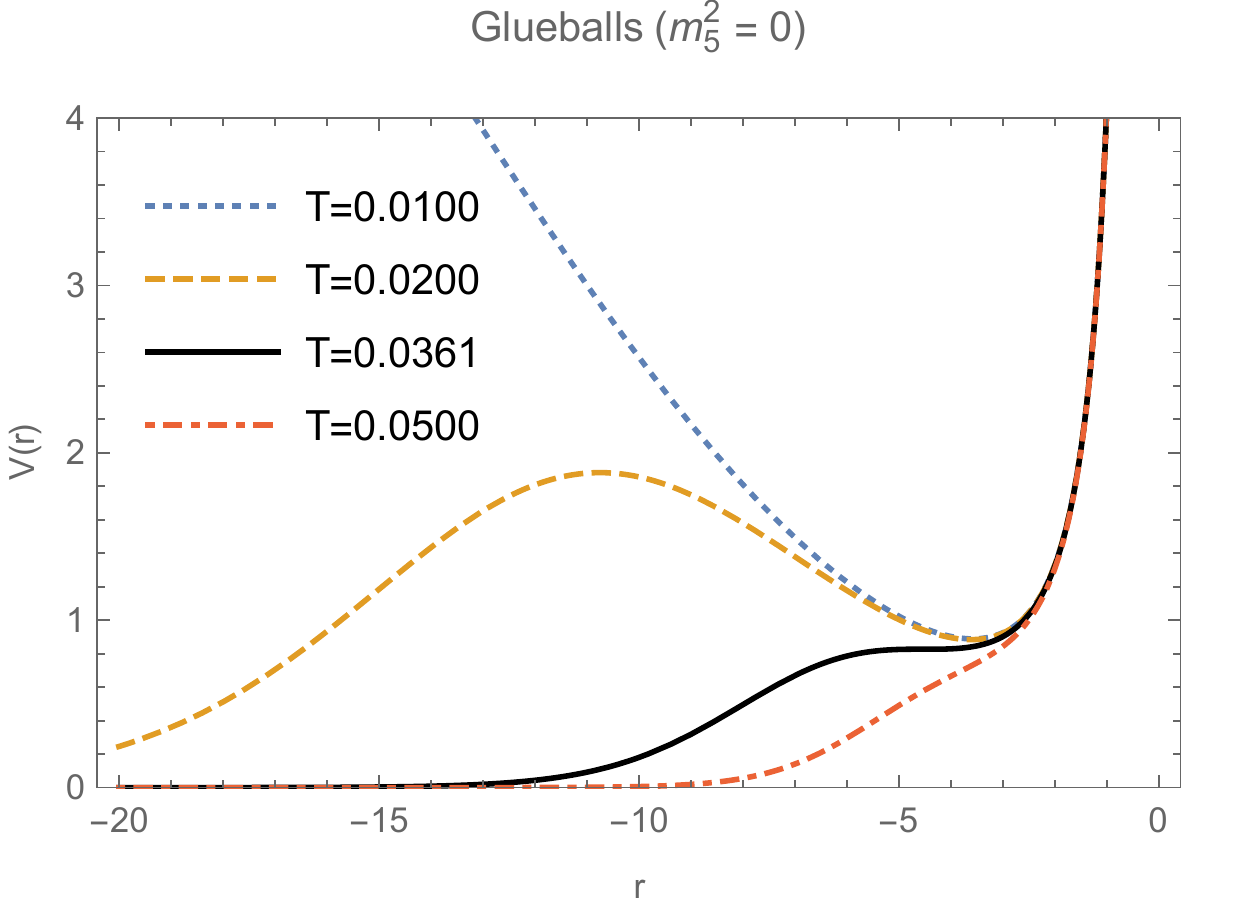}\\
    \includegraphics[width=3.1 in]{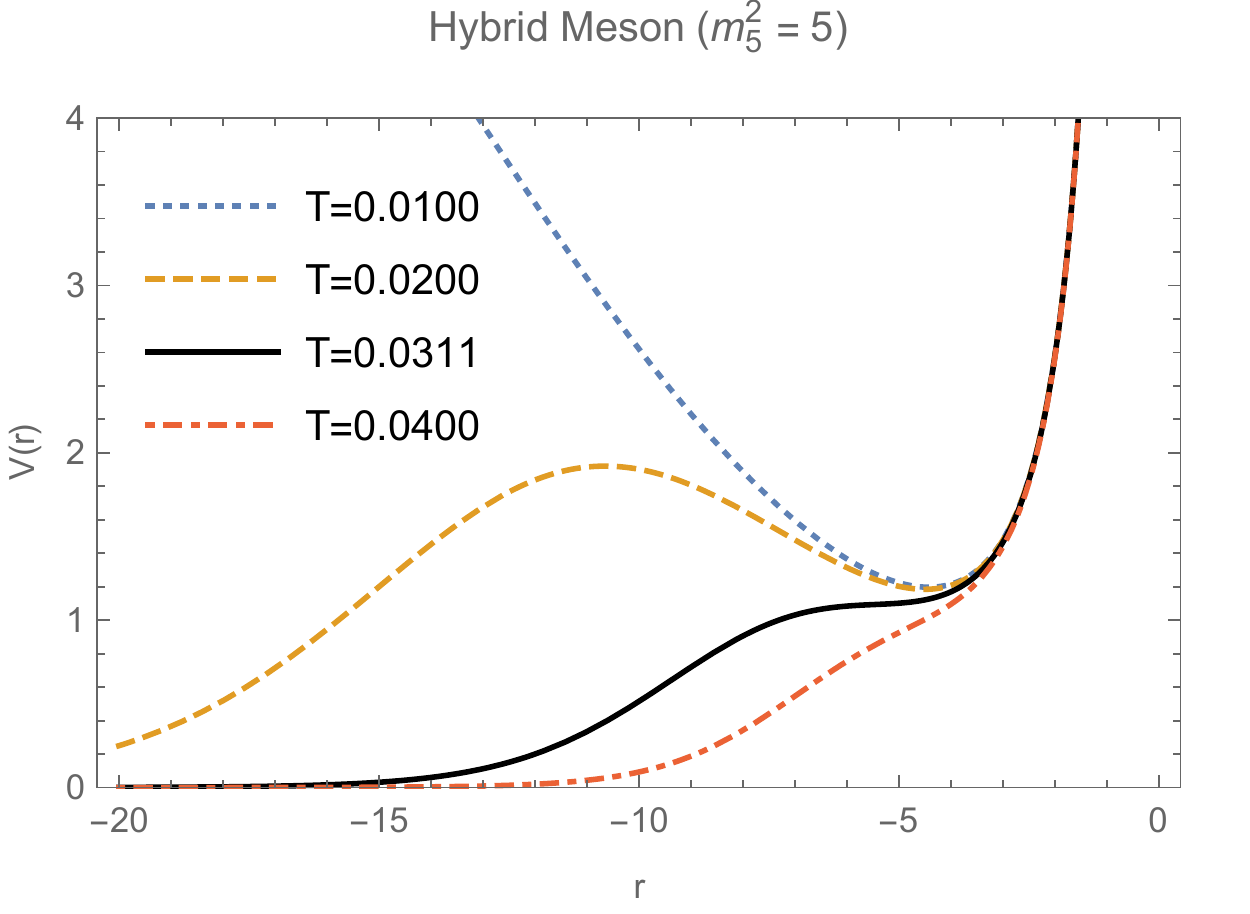}
    \includegraphics[width=3.1 in]{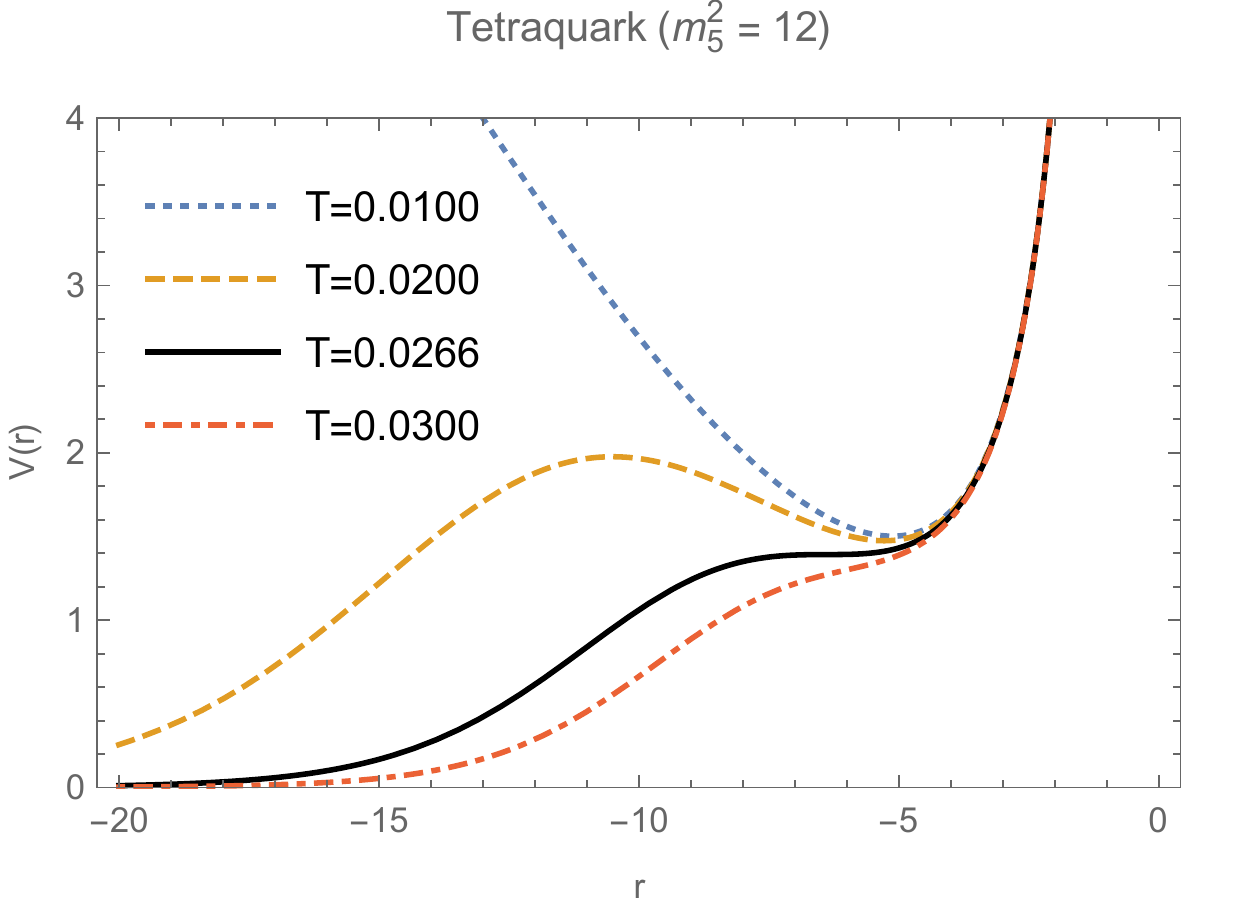}    
  \end{tabular}
\caption{Plots show changes in potential at different temperatures for some light scalar hadrons with a different number of constituents. Continuous lines correspond to cases for the melting temperature for each hadron. Without loss of generality we use $R = 1$. Temperatures are in GeV.}
\end{figure*}
\end{center}

\section{The model}

In AdS / QCD models, scalar hadrons on the AdS side are modelled by scalar modes in a 5D AdS curved space with dilaton. In such models, the AdS mass ($m_{5}$) is related to the dimension of operator that creates hadrons on the gauge theory side, so as this mass takes precise values depending on the hadron considered, this model can be used to study melting temperature for scalar hadrons with an arbitrary number of constituents.

The action considered is (e.g., see \cite{Vega:2008af, Gutsche:2011vb, Vega:2016gip, Colangelo:2008us})

\begin{widetext}
\begin{equation}
S=\frac{1}{2K}\int d^{5}x \sqrt{-g}e^{-\phi \left( z\right) } \left[
g^{MN}\partial _{M}X\left( x,z\right) \partial _{N}X\left( x,z\right) + m_{5}^{2}X^{2}\left( x,z\right) \right]
\end{equation}

In order to introduce temperature in this description, it is necessary to consider a 5D AdS black hole metric \cite{Colangelo:2009ra}.

\begin{equation}
ds^{2}=e^{2A\left( z\right) }\left[ -f\left( z\right)
dt^{2}+\sum_{i=1}^{3}\left( dx^{i}\right) ^{2}+\frac{1}{f\left( z\right) }%
dz^{2}\right],
\end{equation}
or
\begin{equation}
g_{MN}=e^{2A\left( z\right) } diag \left( -f\left( z\right) ,1,1,1,
\frac{1}{f\left( z\right) }\right).
\end{equation}

In this case the equation of motion is
\begin{equation}
e^{B\left( z\right) }f\left( z\right) \partial _{z}\left[ e^{-B\left(
z\right) }f\left( z\right) \partial _{z}\psi \right] -f\left( z\right)
e^{2A\left( z\right) }m_{5}^{2}\psi +\omega ^{2}\psi -f\left( z\right)
q^{2}\psi =0,
\end{equation}
where $B(z) = \phi(z) - 3 A(z)$.

For particles at rest ($\vec{q} = \vec{0}$), the last equation changes to

\begin{equation}
\label{EOMespectro}
\small{\partial _{z}\left[ e^{-B\left( z\right) }f\left( z\right) \partial _{z}\psi %
\right] +\left[ \frac{\omega ^{2}}{e^{B\left( z\right) }f\left( z\right) }
-e^{-\phi \left( z\right) +5A\left( z\right) }m_{5}^{2}\right] \psi =0}.
\end{equation}

By using the Liouville substitution (see appendix), it is possible obtain a Schr\"odinger-type equation related to the previous one, the potential of which is

\begin{equation}
V\left( z\left( \xi \right) \right) = e^{B\left( z\left( \xi
\right) \right) }f\left( z\left( \xi \right) \right) e^{-\phi \left( z\left(
\xi \right) \right) +5A\left( z\left( \xi \right) \right) }m_{5}^{2} -\left[
e^{-2B\left( z\left( \xi \right) \right) }\right] ^{-\frac{1}{4}}\frac{d^{2}%
}{d\xi ^{2}}\left( e^{-2B\left( z\left( \xi \right) \right) }\right) ^{\frac{1}{4}}.
\end{equation}
\end{widetext}

This potential at zero temperature ($f(z) = 1$) has normalizable modes (\cite{Vega:2008af, Gutsche:2011vb, Vega:2016gip, Colangelo:2008us}), but at finite temperature is a potential of quasi-bound states, and depth of the well is reduced when the temperature is increased, and finally at a precise temperature the well will disappear, and this temperature is interpreted as the melting temperature for hadrons in a thermal bath \cite{Miranda:2009qp, Bellantuono:2014lra}.

Here we analyze this potential for a metric and dilaton useful for AdS / QCD model. Additionally, we consider different values for $m_{5}$, which according to the AdS / CFT dictionary is related to the dimension of operators that creates scalar hadrons. Expressions that relate AdS mass with operator dimension are \cite{Aharony:1999ti, Vega:2008af}

\begin{equation}
m^{2}_{5} R^{2} = \Delta (\Delta - 4),
\end{equation}
where $R$ is the AdS radii (without loss of generality can be considered as one) and $\Delta$ is the dimension of operator that creates hadrons. In this work  mesons ($q\bar{q};~\Delta=3;~m^{2}_{5}=-3$), glueballs ($gg;~\Delta=4;~m^{2}_{5}=0$), hybrid mesons ($q\bar{q}g;~\Delta=5;~m^{2}_{5}=5$) and tetraquarks ($qq\bar{q}\bar{q};~\Delta=6;~m^{2}_{5}=12$) \cite{Vega:2008af} are considered, where $q$, $\bar{q}$ and $g$ denote a general quark, antiquark and gluon in the valence configuration for the hadrons studied.

In this work
\begin{equation}
f(z) =1-\frac{z^{4}}{z_{h}^{4}}~~~~~,~~~~~0<z<z_{h},
\end{equation}
was used and therefore,

\begin{equation}
\small{\xi =\int_{0}^{z}\frac{1}{1-\frac{t^{4}}{z_{h}^{4}}}dt=\frac{z_{h}}{2}\left(
-\arctan \left( \frac{z}{z_{h}}\right) +\frac{1}{2}\log \left( \frac{z_{h}-z%
}{z_{h}+z}\right) \right)}
\end{equation}
where  $z_{h}$ is the event horizon, which is related to temperature by 

\begin{equation}
z_{h}=\frac{1}{\pi T}
\end{equation}

To end this section, we wish to mention that in the literature it is possible to find a different approach to calculate melting temperature from an analysis of the potential \cite{Fujita:2009ca}. In this approach, the authors used $u(z) = \sqrt{e^{-B(z)} f(z)} \psi(z)$ in order to transform (\ref{EOMespectro}) into an equation like $-u''(z) + U(z) u(z) = 0$, the potential of which is analyzed similarly to what we discussed previously. This potential depends on the hadron mass also; therefore, it is neccesary to know this mass in the temperature range considered or to make a reasonable supposition about its value. In \cite{Fujita:2009ca}, the $\omega$ value was considered at zero temperature. This approach, although more simple than we used here, only produces an approximation for melting temperature, because it does not consider that hadron masses change with temperature. 

\begin{center}
\begin{figure*}
  \begin{tabular}{c c}
    \includegraphics[width=3.1 in]{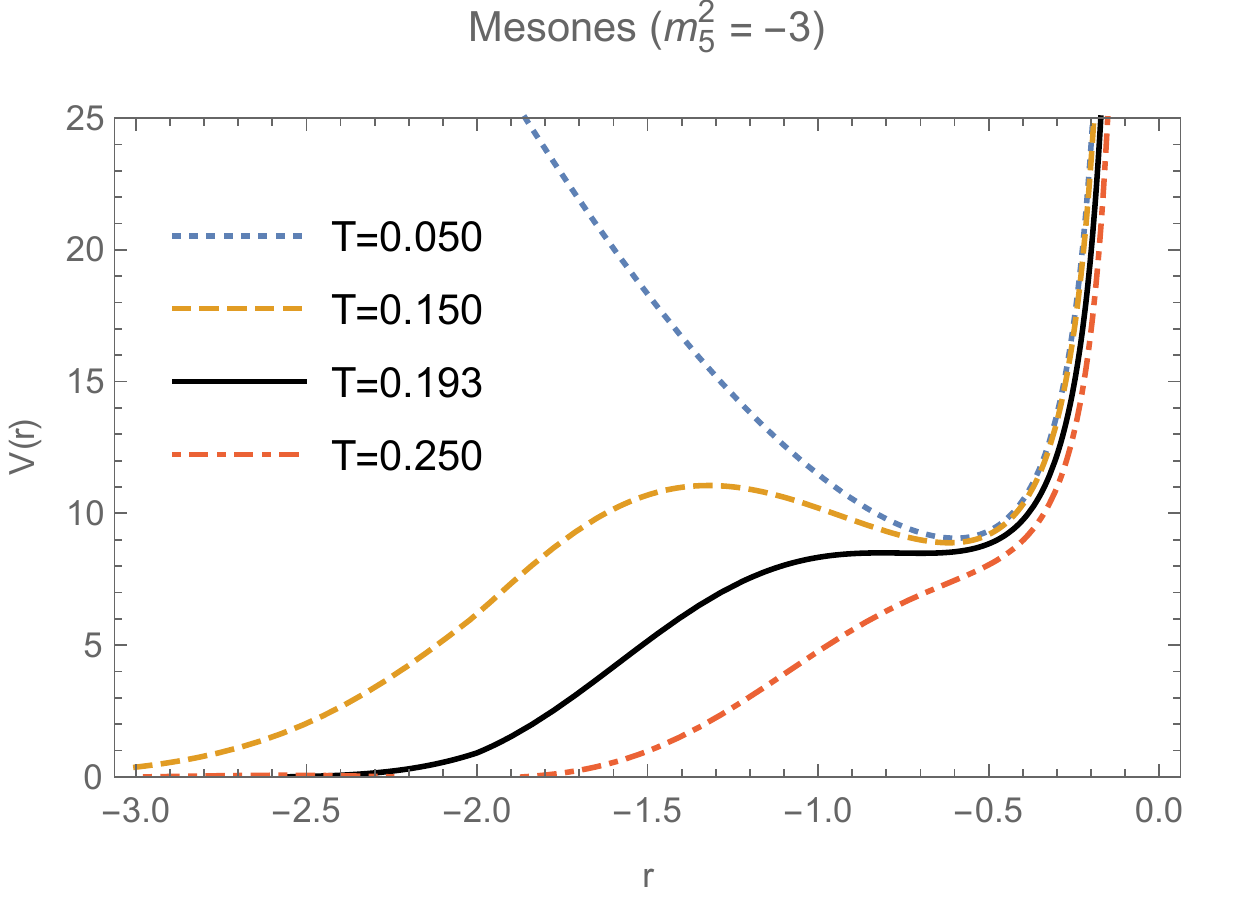}
    \includegraphics[width=3.1 in]{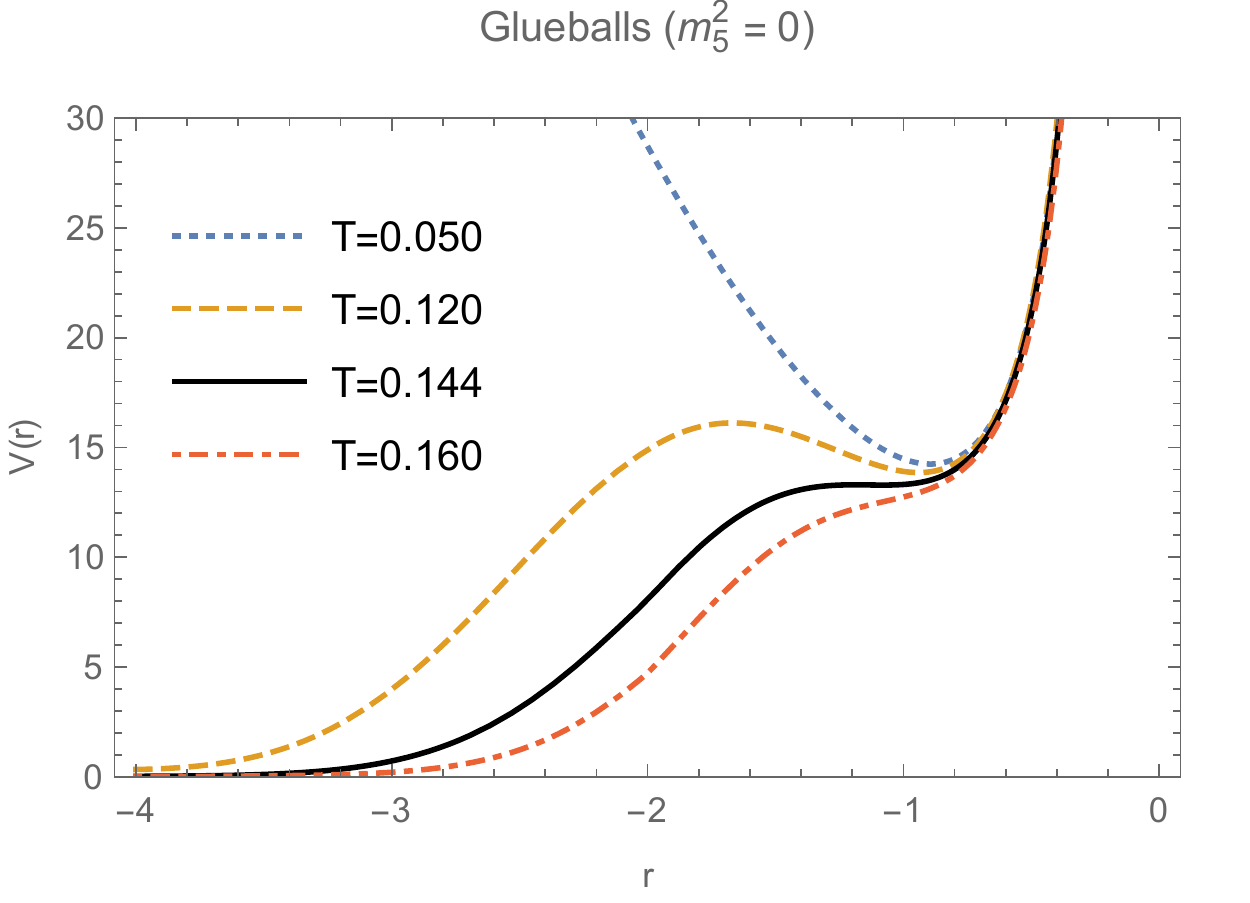}\\
    \includegraphics[width=3.1 in]{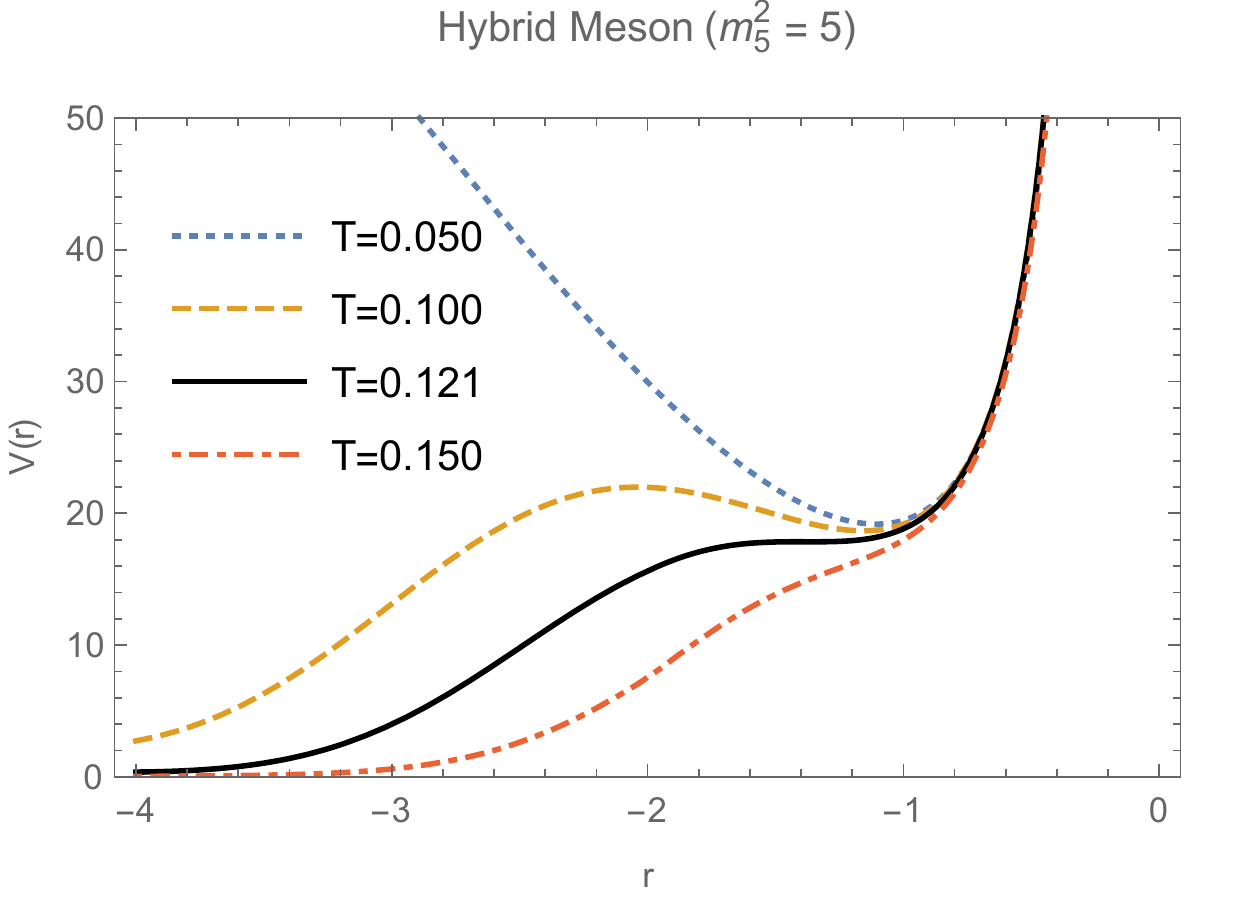}
    \includegraphics[width=3.1 in]{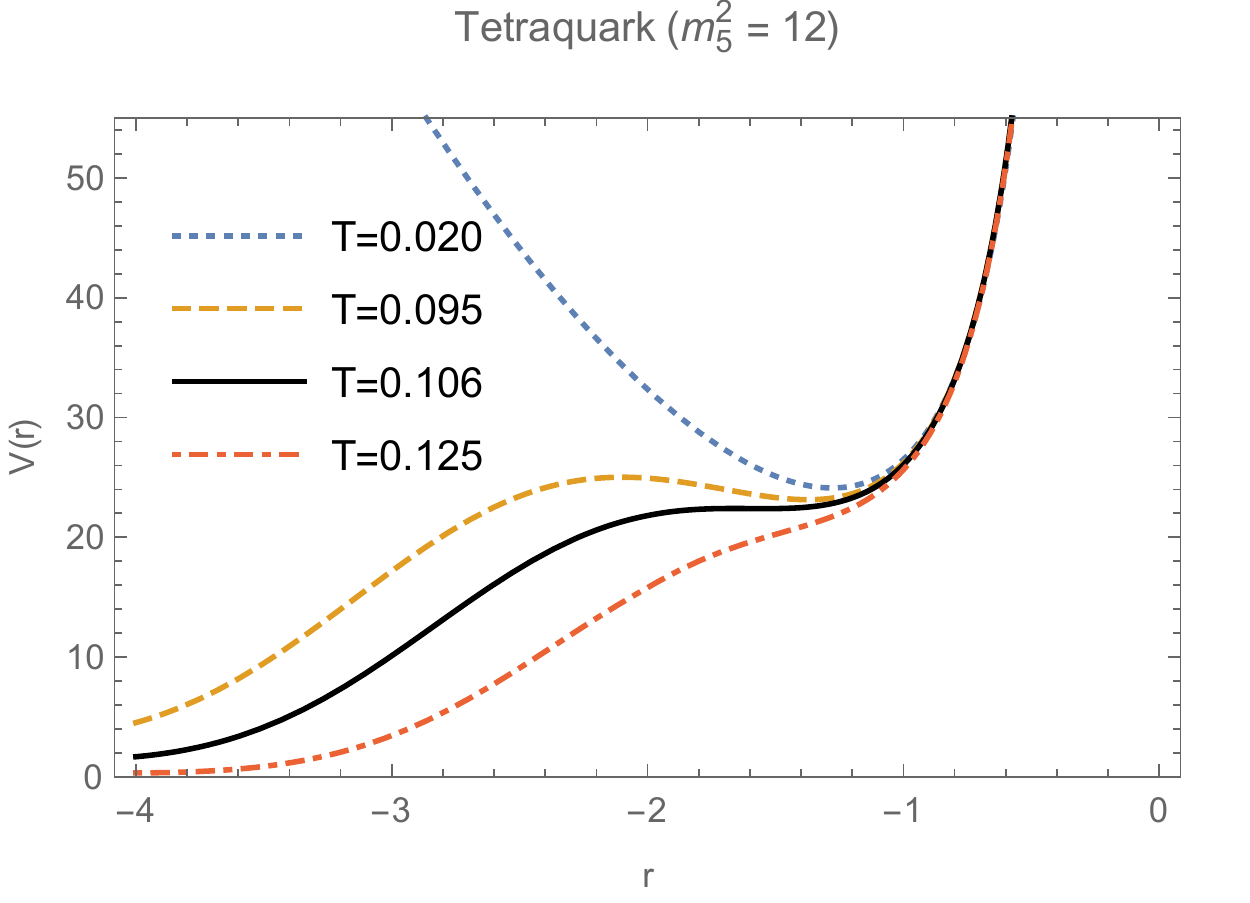}    
  \end{tabular}
\caption{Plots show changes in potential at different temperatures for some heavy scalar hadrons with a different number of constituents. Continuous lines correspond to cases for the melting temperature for each hadron. Without loss of generality we use $R = 1$. Temperatures are in GeV.}
\end{figure*}
\end{center}



\section{Results and Conclusions}

At zero temperature we obtained the traditional soft wall model, and as it is possible to see in plots, an increase in temperature changes the potential and at a characteristic temperature wells in the potential disappear \cite{Miranda:2009qp, Bellantuono:2014lra, Fujita:2009ca}. This characteristic temperature is interpreted as the melting temperature for each hadron in a thermal bath. In the plots presented, the continuous line shows the potential at melting temperature, and as can be see, mesons are the most resistant hadrons to be melted in a thermal bath, and this temperature is increased in the case which considers heavy hadrons. Although we limit ourselves to four kinds of scalar hadrons here, in general when the number of valence constituents is increased, the melting temperature is reduced, so we can say that at a finite temperature it is more difficult to find exotic scalar hadrons, and in general the most resistant scalars are mesons. On the other hand, we found which melting temperature increases when $\kappa$ is increased, and as this parameter in the model considered must be increased in order to include hadrons with heavy quarks in the holographic model, we can conclude that in general heavy scalars are more resistant to being melt in a thermal bath. The idea of mesons with heavy quarks being more resistant to being melted in a thermal bath was introduced several years ago, and today charmonia and bottomonia suppressions are well supported. 

In our opinion, one of the most interesting things discussed here is that it is possible to study hadrons simply with a different number of constituents and ascertain which one is the most resistant at high temperatures. 

This approach can be extended to include different hadrons (spin $1/2$, $1$, $3/2$, etc.) and with small changes could be used to study properties of hadrons in dense media. We intend to cover these topics in future works.



\appendix
\section{Liouville substitution}

Let's consider the equation (e.g., see \cite{Arfken})
\begin{equation}
\frac{d}{dz}\left[ p\left( z\right) \frac{d}{dz}\psi \left( z\right) \right]
+\left[ \omega ^{2}g\left( z\right) -q\left( z\right) \right] \psi \left(
z\right) = 0.
\end{equation}

The Liouville substitution

\begin{equation}
\psi(z) = v (\xi) [p (z) g (z)] ^{-\frac{1}{4}} \xi = \int_{0}^{z} [ \frac{g(t) }{p(t)}]^{\frac{1}{2}}dt
\end{equation}

allows us to get a Schr\"odinger-type equation as,

\begin{equation}
\frac{d^{2}v\left( \xi \right) }{d\xi ^{2}}+\left[ \omega ^{2}-Q\left( \xi
\right) \right] v\left( \xi \right) =0
\end{equation}

where

\begin{equation}
Q\left( \xi \right) =\frac{q\left( z\left( \xi \right) \right) }{g\left(
z\left( \xi \right) \right) } +\left[ p\left( z\left( \xi \right) \right)
g\left( z\left( \xi \right) \right) \right] ^{-\frac{1}{4}}\frac{d^{2}}{d\xi
^{2}}\left( p\left( z\left( \xi \right) \right) g\left( z\left( \xi \right)
\right) \right) ^{\frac{1}{4}} 
\end{equation}



\begin{acknowledgments}
This work was supported by FONDECYT (Chile) under Grant No. 1141280.
\end{acknowledgments}




\end{document}